\documentclass[11pt]{article}
\usepackage{amsmath}
\usepackage{amsfonts}
\usepackage{amssymb}
\usepackage{graphicx}

\def\bea{\begin{eqnarray}}
\def\eea{\end{eqnarray}}
\title{  GMG/GCA description  }

\date{}

\begin{document}
\begin{center}
\LARGE { \bf   Generalized Massive Gravity and Galilean Conformal
Algebra in two dimensions }
\end{center}
\begin{center}
{\bf M.R Setare\footnote{rezakord@ipm.ir} \\  V. Kamali\footnote{vkamali1362@gmail.com}}\\
 {\ Department of Campus of Bijar,  University of Kurdistan,  \\
Bijar, Iran.}
 \\
 \end{center}
\vskip 3cm

{\bf{Abstract:}} Galilean conformal algebra (GCA) in two
dimensions arises as contraction of two copies of the centrally
extended Virasoro algebra ($t\rightarrow t, x\rightarrow\epsilon
x$ with $\epsilon\rightarrow 0$). The central charges of GCA can
be expressed in term of Virasoro central charges. For finite and
non-zero  GCA central charges, the Virasoro central charges must
behave as asymmetric form $O(1)\pm O(\frac{1}{\epsilon})$. We
propose that, the bulk description  for 2d GCA with asymmetric
central charges is given by general massive gravity (GMG) in
three dimensions. It can be seen that, if the gravitational
Chern-Simons coupling $\frac{1}{\mu}$ behaves as of order
O($\frac{1}{\epsilon}$) or ($\mu\rightarrow\epsilon\mu$), the
central charges of GMG have the above $\epsilon$ dependence. So,
in non-relativistic  scaling limit $\mu\rightarrow\epsilon\mu$,
we calculated GCA parameters and finite entropy in term of
gravity parameters mass and angular momentum of GMG.

\newpage
\section{Introduction}
The AdS/CFT correspondence \cite{a1}, has been considered in
during the past decade. Recently, the non-relativistic version of
the duality between gravity theory and boundary quantum field
theory has been attracted a lot of attention. One of the
non-relativistic conformal symmetries  is the schr\"{o}dinger
symmetry group \cite{a2}. This group is the largest symmetry of
the schr\"{o}dinger equation (without potential term), and it can
be seen in cold atoms system \cite{a3}. The gravity dual for
field theory with this symmetry was proposed in \cite{a4}. In this
paper, we would consider another version of non-relativistic
conformal group, named  Galilean conformal (GC) group, which
arises as parametric contraction of conformal group
($t\rightarrow t$, $x_i\rightarrow\epsilon x_i$ with
$\epsilon\rightarrow 0$). Unlike the schr\"{o}dinger group GC
group can be given an infinite extension in any space-time
dimensions. The generators of this group are : ~$L^n=-(n+1)t^n
x_i\partial_i-t^{n+1}\partial_t$,~$M^{n}_i=t^{n+1}\partial_i$
and~$J_{ij}^n=-t^n(x_i\partial_j-x_j\partial_i)$~ for an arbitrary
integer $n$, where  i and  j are specified by the spatial
directions \cite{a5}. Galilean conformal algebra (GCA) was
explored in two dimensions  by two copies  of Virasoro algebra in
non-relativistic limit \cite{1}. In quantum level, 2d CFT was
constructed by two copies of centrally extended Virasoro algebra,
so in non-relativistic limit, we can also consider the quantum
aspects of the centrally extended GCA. The central charges of GCA
($C_1$, $C_2$) can be expressed in term of CFT central charges
($c_L$, $c_R$), and it is known  that GCA central charges is
asymmetric \cite{1}. For finite and non-zero ($C_1$, $C_2$) the
Virasoro central charges must behave as $c_{L}\sim
O(1)+O(\frac{1}{\epsilon})$ and $c_{R}\sim
O(1)-O(\frac{1}{\epsilon})$.\\
In this paper we propose the gravity dual  for 2d GCA. The gravity
dual of the GCA was given in term of Newton-Cartan like
$AdS_2\times R^d$ \cite{a5}. One option for bulk description  of
2d GCA, with asymmetric central charges would be general massive
gravity (GMG) in three dimensions \cite{2, a6}. Massive gravity in
3d was initiated by topological massive gravity (TMG) \cite{a7},
where the action of this theory is realized by the usual
Einstein-Hilbert (EH) term, which includes the cosmological
constant and parity-violating Chern-Simons term, with coupling
$\frac{1}{\mu}$. Massive gravity in three dimensions was extended
by a new kind of three dimensional massive gravity (NMG) \cite{2,
a6}. NMG is defined by adding higher derivative term to the EH
term in action, with coupling $\frac{1}{m^2}$. Finally, in
general model (GMG) the action contains, EH term, the
gravitational CS term and higher derivative terms of NMG. The
general massive gravity central charges are asymmetric
($c_{L,R}=\frac{3l}{2G}(1-\frac{1}{2m^2l^2}\pm\frac{1}{\mu l})$).
 From these central charges, in non-relativistic limit ($\mu\rightarrow\epsilon\mu$), the GCA
 central charges $C_1$ and $C_2$ are  finite. In
 the following  we see that the central charges of GCA, the scaling dimension $\Delta$
 and rapidity $\xi$ (which are the eigenvalues of $L_0$ and $M_0$)
 in non-relativistic limit ($\mu\rightarrow\epsilon \mu, J\rightarrow\epsilon J$)
 are finite and expressed by gravity parameters (M, J, $\mu$,
 $m^2$). Also, the entropy in the non-relativistic limit ($M\rightarrow M$, $J\rightarrow\epsilon J$, $\mu\rightarrow\epsilon\mu$)
 is finite and can be expressed in term of GCA parameters ($C_1, C_2, \Delta,
 \xi$). The rest of the paper is organized as: in section 2 we
 give a brief review of 2d CFT and its contraction, the GCA parameters were realized in term of CFT parameters. In section 3
 we propose general massive gravity is gravity dual of 2d GCA in non-relativistic limit ($\mu\rightarrow\epsilon\mu$), in
 this section GCA parameters were constructed in term of gravity
 parameters and finally we obtained finite entropy in
 non-relativistic limit, and the last section is devoted to the
 conclusion.

\section{GCA in 2d}
Galilean conformal algebra in 2d can be obtained  from
contracting 2d conformal symmetry \cite{1}. 2$d$ Conformal algebra
at the quantum level is described by two copies of centrally
extended  Virasoro algebra. In two dimensions space-time ($z=x+t$,
$\overline{z}=x-t$), the CFT generators
\begin{eqnarray}\label{1}
\mathcal{L}_{n}=-z^{n+1}\partial_{z},
~~~~~~~~~~~~~~~~~~~~~~~~\overline{\mathcal{L}}_n=-\overline{z}^{n+1}\partial_{\overline{z}},
\end{eqnarray}
obey the centrally extended Virasoro algebra
\begin{eqnarray}\label{2}
[\mathcal{L}_m,\mathcal{L}_n]=(m-n)\mathcal{L}_{m+n}+\frac{c_{R}}{12}m(m^2-1)\delta_{m+n,0}, \\
\nonumber
[\overline{\mathcal{L}}_m,\overline{\mathcal{L}}_n]=(m-n)\overline{\mathcal{L}}_{m+n}+\frac{c_{L}}{12}m(m^2-1)\delta_{m+n,0}.
\end{eqnarray}
By taking the non-relativistic limit ($t\rightarrow t$,
$x\rightarrow\epsilon x$ with $\epsilon\rightarrow 0$),  the GCA
generators $L_{n}$ and $M_n$ are constructed from Virasoro
generators by
\begin{eqnarray}\label{3}
L_n=\lim_{\epsilon\rightarrow
0}(\mathcal{L}_n+\overline{\mathcal{L}}_n)=-(n+1)t^n\partial_{x}-t^{n+1}\partial_x, ~~~\\
\nonumber M_{n}=\lim_{\epsilon \rightarrow 0 }\epsilon
(\mathcal{L}_n-\overline{\mathcal{L}}_n)=t^{n+1}\partial_{x}.
~~~~~~~~~~~~~~~~~~~~~
\end{eqnarray}
 From Eqs.(\ref{2}) and (\ref{3}), one obtains centrally extended 2d GCA
\begin{eqnarray}\label{4}
[L_m,L_n]=(m-n)L_{m+n}+C_1m(m^2-1)\delta_{m+n,0}, ~\\
\nonumber [L_m,L_n]=(m-n)M_{m+n}+C_2m(m^2-1)\delta_{m+n,0}, \\
\nonumber [M_n,M_m]=0.
~~~~~~~~~~~~~~~~~~~~~~~~~~~~~~~~~~~~~~~~~~~~~
\end{eqnarray}
The GCA central charges ($C_1$, $C_2$) are related to CFT central
charges ($c_L$, $c_R$) as:
\begin{eqnarray}\label{5}
C_1=\lim_{\epsilon\rightarrow 0}\frac{c_L+c_R}{12},
~~~~~~C_2=\lim_{\epsilon\rightarrow
0}(\epsilon\frac{c_L-c_R}{12}).
\end{eqnarray}
From above equations, for a non-zero  and finite ($C_2$, $C_1$) in
the limit $\epsilon\rightarrow 0$, it can be seen that we need
$c_L-c_R\propto O(\frac{1}{\epsilon})$ and $c_L+c_R \propto
O(1)$.  Similarly, rapidity $\xi$ and scaling dimensions
$\Delta$, which are the eigenvalues of $M_0$ and $L_0$
respectively, are given by
\begin{eqnarray}\label{6}
\Delta=\lim_{\epsilon \rightarrow 0}(h+\overline{h}),
~~~~~~~~~\xi=\lim_{\epsilon\rightarrow 0}\epsilon
(h-\overline{h}),
\end{eqnarray}
where $h$ and $\overline{h}$ are eigenvalues of $\mathcal{L}_0$
and $\overline{\mathcal{L}}_0$ respectively. Equation (\ref{6})
tells us that, $h+\overline{h}$ is of order $O(1)$ while
$h-\overline{h}$ must be order $O(\frac{1}{\epsilon})$, for the
finite $\Delta$, $\xi$.

\section{GCA realization of general massive gravity }
In this section we would like to propose that the contracted BTZ
black hole solution of  three dimensional general massive gravity
(GMG) is gravity dual of 2$d$ GCA in the context of $AdS/CFT$
correspondence. It is notable that the GMG (as a gravity dual)
has to yield finite parameters ($\Delta, \xi, C_1, C_2$ and
entropy $S_{GCA}$) for GCA. The action of the cosmological
general massive gravity in three dimensions is \cite{2}
\begin{eqnarray}\label{7}
S[g_{\mu\nu}]=\frac{1}{16\pi G}\int \sqrt{-g}(R-2\lambda
m^2+\frac{1}{m^2}\mathcal{L}_{NMG}+\frac{1}{\mu}\mathcal{L}_{CS})d^3x,
\end{eqnarray}
where the NMG term is
\begin{eqnarray}\label{8}
\mathcal{L}_{NMG}=R_{\mu\nu}R^{\mu\nu}-\frac{3}{8}R^2,
\end{eqnarray}
and the gravitational Chern-Simons (CS) term is
\begin{eqnarray}\label{9}
\mathcal{L}_{CS}=\frac{1}{2} \varepsilon^{\mu\nu\rho}
(\Gamma^{\alpha}_{\mu\beta}\partial_{\nu}\Gamma^{\beta}_{\rho\alpha}+
\frac{2}{3}\Gamma^{\alpha}_{\mu\beta}\Gamma^{\beta}_{\nu\gamma}\Gamma^{\gamma}_{\rho\alpha}).
\end{eqnarray}
The Einstein equation of motion of this action is
\begin{eqnarray}\label{10}
G_{\mu\nu}+\lambda
m^2g_{\mu\nu}+\frac{1}{2m^2}K_{\mu\nu}+\frac{1}{\mu}C_{\mu\nu}=0,
\end{eqnarray}
where $G_{\mu\nu}$ is the Einstein tensor,  the  tensor
$C_{\mu\nu}$ (Cotton tensor), due to the gravitational CS term,
reads
\begin{eqnarray}\label{11}
C_{\mu\nu}=\varepsilon_{\mu}^{\alpha\rho}D_{\alpha}(R_{\beta\nu}-\frac{1}{4}Rg_{\beta\nu}),
~~~~\varepsilon^{012}=1,
\end{eqnarray}
and the tensor $K_{\mu\nu}$, coming from  NMG term
\begin{eqnarray}\label{12}
K_{\mu\nu}=2D^2R_{\mu\nu}-\frac{1}{2}(D_{\mu}D_{\nu}R+g_{\mu\nu}D^2R)-8R_{\mu\alpha}R^{\alpha}_{\nu}\\
\nonumber
+\frac{9}{2}RR_{\mu\nu}+(3R^{\alpha\beta}R_{\alpha\beta}-\frac{13}{8}R^2)g_{\mu\nu}.
~~~~~~~~~~~~~~~~
\end{eqnarray}
The reason of this choice of $\mathcal{L}_{NMG}$ is that
$g^{\mu\nu}K_{\mu\nu}=\mathcal{L}_{NMG}$. The parameter $\lambda$
is dimensionless and characterizes the cosmological constant
term, while $m$ has the dimension of mass and provides the
coupling to the NMG term, also $CS$ coupling $\mu$ has the
dimension of mass. The solution of BTZ black hole is given by
\begin{eqnarray}\label{13}
ds^2=(-f(r)+\frac{16G^2J^2}{r^2})dt^2+\frac{dr^2}{f(r)}+r^2d\varphi^2+8GJdtd\varphi,
\end{eqnarray}
where
\begin{eqnarray}\label{14}
f(r)=(\frac{r^2}{l^2}-8GM+\frac{16G^2J^2}{r^2}). ~~~~~~~
\end{eqnarray}
The parameters $M$ and $J$ correspond to the mass and angular
momentum in the case without the terms  $\mathcal{L}_{NMG}$ and
$\mathcal{L}_{CS}$ , but their definitions in the case with these
terms are \cite{a}
\begin{eqnarray}\label{15}
M_1=(1-\frac{1}{2m^2l^2})(M+\frac{1}{\mu}\frac{J}{l^2}),
~~~~~~~~~~J_1=(1-\frac{1}{2m^2l^2})(J+\frac{1}{\mu}M).
\end{eqnarray}
Due to NMG and CS terms, the Bekenstein-Hawking entropy is given
by
\begin{eqnarray}\label{16}
S_{BH}=\frac{\pi r_+}{2G}(1-\frac{1}{2m^2l^2})+\frac{1}{\mu
l}\frac{\pi r_-}{2G},
\end{eqnarray}
where horizons $r_{\pm}$ are defined by
\begin{eqnarray}\label{17}
r_{\pm}=\sqrt{2Gl(lM+J)}\pm\sqrt{2Gl(l M-J)}.
\end{eqnarray}
In GMG case the parity-violating Chern-Simons terms leads to
different   central charges of the Virasoro algebra.
\begin{eqnarray}\label{18}
c_L=\frac{3l}{2G}(1-\frac{1}{2m^2l^2}+\frac{1}{\mu l}),
~~~~~~~~~~c_R=\frac{3l}{2G}(1-\frac{1}{2m^2l^2}-\frac{1}{\mu l}).
\end{eqnarray}
For the BTZ black hole solution in GMG, $h$ and $\overline{h}$ are
calculated as
\begin{eqnarray}\label{19}
h=\frac{1}{2}(lM_1+J_1)+\frac{c_L}{24},
~~~~~~~~\overline{h}=\frac{1}{2}(l M_1-J_1)+\frac{c_R}{24}.
\end{eqnarray}
Then, the microscopic entropy is expressed by Cardy formula (in
$1\leq\mu l$ case.)
\begin{eqnarray}\label{20}
S_{CFT}=2\pi(\sqrt{\frac{c_L
h}{6}}+\sqrt{\frac{c_R\overline{h}}{6}})=\frac{\pi
r_+}{2G}(1-\frac{1}{2m^2l^2})+\frac{1}{\mu l}\frac{\pi r_-}{2G},
\end{eqnarray}
which is in good agreement  with Bekenstein-Hawking entropy. Now
we consider non-relativistic limit in three dimensional general
massive  gravity.
\begin{eqnarray}\label{21}
t\rightarrow t, ~~~~~~~r\rightarrow r,
~~~~~~~\varphi\rightarrow\epsilon \varphi.
\end{eqnarray}
The parameters $M$ and $J$ in the BTZ solution should scale like
\begin{eqnarray}\label{22}
M\rightarrow M, ~~~~~~~ J\rightarrow \epsilon J.
\end{eqnarray}
It is seen in (\ref{5}), for non-zero and finite GCA central
charges ($C_1$, $C_2$), the Virasoro central charges must behave
as $c_L+c_R\sim O(1)$ and $c_L-c_R\sim O(\frac{1}{\epsilon})$. In
GMG case using Eq.(\ref{8}) we have  $c_L-c_R=\frac{3}{G\mu }$,
so in non-relativistic limit we demand that $\mu$ should scales as
\begin{eqnarray}\label{23}
\mu\rightarrow \epsilon\mu.
\end{eqnarray}
From Eqs.(\ref{5}), (\ref{18}),  and (\ref{23}) the GCA central
charges $C_1$ and $C_2$  are finite
\begin{eqnarray}\label{24}
C_1=\frac{l}{4G}(1-\frac{1}{2m^2l^2}), ~~~~~~C_2=\frac{1}{4G\mu}.
\end{eqnarray}
Similarly, from Eqs.(\ref{6}), (\ref{19}) and (\ref{23}), scaling
dimensions $\Delta$ and rapidity $\xi$, which are the eigenvalue
of $L_0$ and $M_0$ are given by
\begin{eqnarray}\label{25}
\Delta=\lim_{\epsilon\rightarrow 0} (l
M_1+\frac{c_L+c_R}{24})=(1-\frac{1}{2m^2l^2})(lM+\frac{1}{\mu l}J)+\frac{C_1}{2}, \\
\nonumber \xi=\lim_{\epsilon\rightarrow
0}\epsilon(J_1+\frac{c_L-c_R}{24})=(1-\frac{1}{2m^2l^2})(\frac{1}{\mu}M)+\frac{C_2}{2}.
~~~~~~~~
\end{eqnarray}
In the above equations, when $M$ and $J$ are large enough, the
terms $\frac{C_1}{2}$ and $\frac{C_2}{2}$ can be neglected. The
dual theory (GMG) of GCA in non-relativistic limit (\ref{23}) has
to yield the finite ($C_1$, $C_2$) (\ref{24}) and finite
($\Delta$, $\xi$) (\ref{25}). In the following, we  would like to
obtain the finite entropy of the GCA. The scaling limit
($M\rightarrow M$, $J\rightarrow \epsilon J$) requires that the
event horizons of the BTZ black hole should scale
\begin{eqnarray}\label{26}
r_+\rightarrow 2l\sqrt{2GM}, ~~~~~~~~~~~~r_-\rightarrow \epsilon
\sqrt{\frac{2G}{M}}J,
\end{eqnarray}
so the black hole  entropy (\ref{16}) is given by
\begin{eqnarray}\label{27}
\lim_{\epsilon \rightarrow 0 }S_{BH}=\lim_{\epsilon \rightarrow 0
}S_{CFT}=\pi(1-\frac{1}{2m^2l^2})\sqrt{\frac{2l^2M}{G}}+\frac{J}{\mu
l }\sqrt{\frac{1}{2GM}}.
\end{eqnarray}
From the expressions of the central charges $C_{1}$ and $C_2$
(\ref{24}), scaling dimension $\Delta$ and rapidity $\xi$
(\ref{25}), we can rewrite the finite entropy as
\begin{eqnarray}\label{28}
S_{GCA}=\pi(2C_1\sqrt{\frac{2\xi}{C_2(1-\frac{1}{2m^2l^2})}}+\Delta
\sqrt{\frac{2C_2}{\xi(1-\frac{1}{2m^2l^2})}}-\\
\nonumber
(1-\frac{1}{2m^2l^2})^{-1}C_1\sqrt{\frac{2\xi}{C_2(1-\frac{1}{2m^2l^2})}}
). ~~~~~~~~~~~~~~~~~~~~~~
\end{eqnarray}
This expression is the entropy for the GCA in two dimensions. One
can see that the relation (\ref{28}) in CTMG limit
($\frac{1}{m^2}\rightarrow 0$) is agree with the result of
\cite{3}.

\section {Conclusion}
Recently the authors of \cite{3} (see also \cite{bag1}) have shown
that the $GCA_2$ is the asymptotic symmetry of Cosmological
Topologically Massive Gravity (CTMG) in the non-relativistic
limit. They have obtained the central charges of $GCA_2,$ and
also a non-relativistic generalization of Cardy formula. Following
this work, in this paper we propose the general massive gravity
(GMG) as a gravity dual of $2d$ GCA in the context of the
non-relativistic $AdS_3/CFT_2$ correspondence. At the quantum
level the centrally extended GCA arises  precisely from a
non-relativistic contraction of the two copies of the Virasoro
algebra. The relations between GCA and CFT central charges
(\ref{5}), tell us that, for non-zero and finite GCA central
charges ($C_1$, $C_2$), the CFT central charges ($c_{L, R}$)
should behave as asymmetric form $O(1)\pm O(\frac{1}{\epsilon})$
with $\epsilon\rightarrow 0$. In GMG  the parity violating
Chern-Simons term leads to left-right asymmetric central charges
(\ref{18}). It is notable that the gravity dual in
non-relativistic limit has to yield finite GCA central charges.
To implement this we propose $\mu\rightarrow \epsilon \mu$, so
the central charges of parent $CFT_2$ which has the above
$\epsilon$ dependence. The non-relativistic limit on the gravity
dual is given by (\ref{21}), (\ref{22}) and (\ref{23}). The GCA
parameters ($C_1$, $C_2$) and ($\Delta, \xi$) in terms of gravity
parameters are given by equations (\ref{24}) and (\ref{25})
respectively. Finally, we calculated finite entropy of the
$GCA_2$, which reduce to the result of \cite{3} in CTMG limit
($\frac{1}{m^2}\rightarrow 0$).

\end{document}